\documentclass[prl,twocolumn,showpacs,superscriptaddress,floatfix]{revtex4}
\usepackage{graphics}  
\usepackage{makeidx}
\usepackage{colordvi}
\RequirePackage{xspace}
\usepackage{relsize}
\def\babar{\mbox{\slshape B\kern-0.1em{\smaller A}\kern-0.1em
    B\kern-0.1em{\smaller A\kern-0.2em R}}}
\def\Dz      {\ensuremath{D^0}\xspace}
\def\Y#1S{\ensuremath{\Upsilon{(#1S)}}\xspace}
\newcommand{\gevc}{\ensuremath{{\mathrm{\,Ge\kern -0.1em V\!/}c}}\xspace}
\newcommand{\mevc}{\ensuremath{{\mathrm{\,Me\kern -0.1em V\!/}c}}\xspace}
\newcommand{\gevcc}{\ensuremath{{\mathrm{\,Ge\kern -0.1em V\!/}c^2}}\xspace}
\newcommand{\mevcc}{\ensuremath{{\mathrm{\,Me\kern -0.1em V\!/}c^2}}\xspace}
\def\Dstarp  {\ensuremath{D^{*+}}\xspace}

\def\pip   {\ensuremath{\pi^+}\xspace}
\def\pim   {\ensuremath{\pi^-}\xspace}
\def\pep2{PEP-II}

\begin{document}  

\begin{flushleft}
\babar-PUB-04/027\\
SLAC-PUB-10594
\end{flushleft}

\title{Search for flavor-changing neutral current and\\
lepton-flavor violating decays of $\mathbf{D^0 \to \ell^+ \ell^-}$}
 
%
\author{B.~Aubert}
\author{R.~Barate}
\author{D.~Boutigny}
\author{F.~Couderc}
\author{J.-M.~Gaillard}
\author{A.~Hicheur}
\author{Y.~Karyotakis}
\author{J.~P.~Lees}
\author{V.~Tisserand}
\author{A.~Zghiche}
\affiliation{Laboratoire de Physique des Particules, F-74941 Annecy-le-Vieux, France }
% ...
\collaboration{The \babar\ Collaboration}
\noaffiliation

\begin{abstract} 
We report on a search for the flavor-changing neutral current decays 
$\Dz\to e^+e^-$ and $\Dz\to\mu^+\mu^-$, 
and the lepton-flavor violating
decay $\Dz\to e^\pm\mu^\mp$. The measurement is based on 
$122\,\mbox{fb}^{-1}$ of data collected by the \babar\ detector 
at the PEP-II asymmetric $e^+e^-$ collider. No evidence is found for
any of the decays. The upper limits on the branching fractions,
at the 90\,\% confidence level, are $1.2\times 10^{-6}$ for
$\Dz\to e^+e^-$, $1.3\times 10^{-6}$ for $\Dz\to\mu^+\mu^-$, and 
$8.1\times 10^{-7}$ for
$\Dz\to e^\pm\mu^\mp$.

\end{abstract}
 
\maketitle    
In the Standard Model (SM), the flavor-changing neutral current (FCNC)
decays $\Dz\to e^+e^-$ and $\Dz\to\mu^+\mu^-$~\cite{conjugate} 
are highly suppressed
by the Glashow-Iliopoulos-Maiani (GIM) mechanism~\cite{Glashow:gm}. 
Their decay 
branching fractions have been estimated to be less than 
$10^{-13}$ even with long-distance processes included.
This prediction is orders of magnitude beyond the
reach of current experiments. Furthermore, the lepton-flavor
violating (LFV) decay $\Dz\to e^\pm\mu^\mp$ is strictly forbidden
in the SM~\cite{neutrino}.

Some extensions to the Standard
Model can enhance the FCNC processes by many orders
of magnitude. 
For example, $R$-parity violating supersymmetry
can increase the branching fractions of 
$\Dz\to e^+e^-$ and $\Dz\to\mu^+\mu^-$ to as high as 
$10^{-10}$ and $10^{-6}$, 
respectively~\cite{Burdman:2001tf}.
The same model also predicts the $\Dz\to e^\pm\mu^\mp$ 
branching fraction
to be of the order of $10^{-6}$. 
The upper bounds on the predicted branching fractions of
$\Dz\to\mu^+\mu^-$ and $\Dz\to e^\pm\mu^\mp$ are close 
to the current experimental sensitivities. As a result, searching for
the FCNC and LFV decays in the charm sector is a potential way 
to test the SM and explore new physics. Similar arguments 
hold for  rare $K$ and $B$ decays, but the charm 
decay is unique since it is sensitive to new physics coupling to
the up-quark sector.

In this paper, we present a search for the decays of
$\Dz\to e^+e^-$, $\Dz\to\mu^+\mu^-$, and
$\Dz\to e^\pm\mu^\mp$. The analysis is based on 
$122\,\mbox{fb}^{-1}$ of data collected on or near the 
\Y4S resonance by the \babar\ detector 
at the PEP-II asymmetric $e^+e^-$ collider.

The \babar\ detector, which is fully described in~\cite{Aubert:2001tu},
provides charged-particle tracking through a combination of a
five-layer double-sided silicon micro-strip detector (SVT) and a
40-layer central drift chamber (DCH), both operating in a
1.5\,T magnetic field in order to provide momentum
measurements. The identification of charged kaons and 
pions is achieved through
measurements of particle energy-loss ($dE/dx$) in the tracking system
and Cherenkov cone angle ($\theta_c$) in a detector of internally
reflected Cherenkov light.  
Electrons are identified primarily in a segmented CsI(Tl) 
electromagnetic calorimeter, while muons are identified
by their penetration through the iron plates of the magnet flux return.

The charmed mesons considered for this analysis originate from 
the fragmentation of charm quarks in the continuum $e^+e^-\to c\bar{c}$
process. There is no advantage in including $D^0$ decays from 
the $B$ mesons because of their higher combinatoric background.
The $\Dz\to\ell^+\ell^-\;(\ell=e,\mu)$ branching ratio is determined by
\begin{equation}
\mathcal{B}(\Dz\to\ell^+\ell^-)=S (N_{\rm obs}-N_{\rm bg}),
\end{equation}
where $N_{\rm obs}$ is the number of $\Dz\to\ell^+\ell^-$ 
candidates observed, $N_{\rm bg}$ is the expected background
and $S$ is the sensitivity factor, defined as:
\begin{equation}
S\equiv\mathcal{B}(\Dz\to\pi^+\pi^-) \frac{1}{N_{\pi\pi}} 
\frac{\epsilon_{\pi\pi}}{\epsilon_{\ell\ell}}.
\end{equation}
Here $\mathcal{B}(\Dz\to\pi^+\pi^-)=(1.43\pm0.07)\times 10^{-3}$ is 
the $\Dz\to\pi^+\pi^-$ 
branching fraction~\cite{Hagiwara:fs}, $N_{\pi\pi}$ is the number
of reconstructed  $\Dz\to\pi^+\pi^-$ decays, $\epsilon_{\ell\ell}$ and
$\epsilon_{\pi\pi}$ are the efficiency for the corresponding
 decay mode. 
We choose $\Dz\to\pi^+\pi^-$ as the normalization mode because 
it is kinematically similar to $\Dz\to\ell^+\ell^-$ and therefore
many common systematic uncertainties cancel in the calculation 
of the efficiency ratio $\epsilon_{\pi\pi}/\epsilon_{\ell\ell}$. 
The key to the analysis is to reduce backgrounds as 
much as possible
while maintaining a high signal efficiency. 

We first outline the general event selection requirements
common to all the data samples used in the analysis and later
describe tighter optimized criteria specific to each decay 
modes.
A pair of oppositely charged tracks is selected to form
a \Dz candidate. They are fit to a common vertex and 
only the candidates with fit probability larger than 1\,\% are 
retained. 
Since charmed mesons from $e^+e^-\to c\bar{c}$ events are 
produced with momenta higher on average than those from 
$e^+e^-\to b\bar{b}$ events,
a minimum value of 2.4\gevc is imposed on the  
center-of-mass momentum of each \Dz candidate.
In order to further reduce the 
background, the \Dz candidate is required to be
from a $\Dstarp\to\Dz\pi^+$ decay. 
The \Dz candidate and the pion from the \Dstarp 
are  fit to a common vertex with a beam 
spot constraint. The probability for this fit is required 
to exceed 1\,\%. The resolution of the mass difference between 
the reconstructed \Dstarp and \Dz candidates is 
approximately 0.25\,\mevcc. We require that 
$|\delta m|\le 2.0\,\mevcc$,
where $\delta m=m(\Dz\pi^+)-m(\Dz)-145.4\mevcc$. 
In addition, all the tracks are required to have 
a minimum number of measurement points in the SVT and the DCH.

We require the electron and muon candidates to have
momenta larger than 0.5\,GeV/$c$ and 1.0\,GeV/$c$ 
in the laboratory frame respectively.
In this range, the average
electron and muon efficiencies are about 95\,\% and 60\,\%, 
and their 
hadron misidentification probabilities are measured 
from $\tau$ decay control samples to be around 0.2\,\%
and 2.0\,\%. Pion identification
is also applied to the daughters of $\Dz\to\pi^+\pi^-$ decays. 
The corresponding single pion identification efficiency 
is around 90\,\%.
No particle identification (PID) is applied on the soft pion from 
the \Dstarp decay.

Except for particle identification, 
the selection criteria applied to
the $\Dz\to\pi^+\pi^-$ mode are the same as those used 
for the $\Dz\to\ell^+\ell^-$ modes.
The signal efficiencies of $\Dz\to\ell^+\ell^-$ and $\Dz\to\pi^+\pi^-$
are evaluated using a Monte Carlo (MC) simulation. 
We use PYTHIA~\cite{Sjostrand:2000wi} for
the fragmentation of the produced $c\bar{c}$.
The final state radiative effects 
are simulated for all decays using PHOTOS~\cite{Barberio:1993qi}. 
The detector response is 
simulated with GEANT4~\cite{Agostinelli:2002hh}, 
and the simulated events are then reconstructed 
in the same manner as the data.

Due to large final-state radiation and bremsstrahlung backgrounds, 
the invariant mass distributions, $m_{\ell\ell}$, of $\Dz\to e^+e^-$ and 
$\Dz\to e^\pm\mu^\mp$ have a low mass tail. We define
an asymmetric signal mass window 
($1.8045\le m_{\ell\ell}\le 1.8845\,\gevcc$) for all three decay
modes. The lower boundary of the signal mass window is chosen to 
include the majority of the radiative
tail of $\Dz\to e^+e^-$ and $\Dz\to e^\pm\mu^\mp$.
The higher boundary corresponds to a little
more than $2\sigma$ of the $\Dz$ mass resolution measured from 
the $\Dz\to\pi^+\pi^-$ control sample. 
In order to avoid any possibility of bias, a blind analysis
technique has been adopted. All events inside the
\Dz mass window  were hidden from inspection until the final 
event selection criteria were established and all systematic
uncertainties were determined.

The $\Dz\to\ell^+\ell^-$ background can be taken as a sum of 
two components: a peaking background
from  $\Dz\to h^+h^-\;(h=K,\pi)$ decays and a 
combinatoric background from other sources. 
The copious two body hadronic $\Dz\to h^+h^-$ decays will mimic
the dilepton signals if both hadrons are misidentified as leptons.
MC studies show that only the decay $\Dz\to\pi^+\pi^-$ 
contributes in the signal window. The $\Dz\to K^-\pi^+$
and $\Dz\to K^+K^-$ backgrounds peak in the lower mass region
because of the high kaon mass.
To estimate the number of peaking background events, $N^{hh}_{\rm bg}$, we
apply the selection criteria for $\Dz\to\ell^+\ell^-$ to
MC simulated $\Dz\to\pi^+\pi^-$ events with lepton 
misidentification rates measured from a control sample.

MC studies show that the combinatoric background in both the 
signal mass window and high mass sideband region 
($1.9045\le m_{\ell\ell}\le 2.0545\,\gevcc$)
is dominated by the combination
of two random leptons. The invariant mass distribution of the 
random lepton pair is flat.  This is indeed
consistent with what is observed in the high mass sideband of the data.
As a result, the expected combinatoric background in the signal
window is just the number of dilepton events in the high mass sideband
scaled by the ratio of the width of the signal region to the high 
mass sideband region. 

To further reduce the background, 
we added a selection on the proper decay time, $ct$, of the
\Dz candidate, and tightened our selections on the
signal mass window and $\delta m$.
We determine the optimal selection criteria by maximizing the 
value $\epsilon_{\ell\ell}/N_{\rm sens}$, 
where $N_{\rm sens}$ is the averaged
90\,\% confidence level upper limit on the number of observed signal 
events that 
would be obtained by an ensemble of experiments with the expected 
background and no real signal~\cite{Feldman:1997qc}. 
Studies show that the correlations
among the optimized discriminating variables are negligible. 

The expected combinatoric background
$N^{\rm comb}_{\rm bg}$ therefore can be factorized as
\begin{equation}
N^{\rm comb}_{\rm bg}=N_{\rm SB} R_{\rm mass} R_{\delta m}
 R_{ct},
\end{equation}
where $N_{\rm SB}$ is the number of high mass sideband events 
passing the 
loose event selection criteria; $R_{\rm mass}$ is the expected 
background rejection factor for a given signal mass window;
$R_{\delta m}$ and $R_{ct}$ are the expected background rejection 
factors for the tighter
$\delta m$ requirement and the $ct$ requirement respectively. 
In order to avoid possible bias due to the statistical fluctuation in 
the high mass sideband, we determine the discriminating 
variable distribution shape
of the combinatoric background from the MC and  $D^0\to\pi^+\pi^-$ 
control sample.
The information is subsequently used to predict the background 
changes as a function of a particular set of selection criteria
rather than by
directly examining the data in the high mass sideband.
The optimized final selection criteria are summarized in 
Table~\ref{table:sel}. The estimated numbers of background events
are listed in Table~\ref{tbl:br}.
\begin{table}[htb]
\begin{center}
\begin{tabular}{cccc}\hline\hline
Mode & $m_{\ell\ell}\;[\gevcc]$  & $\delta m\;[\mevcc]$ & $ct$ \\\hline
$ee$ & $1.8245\le$$m_{ee}$$\le1.8845$ &  
$|\delta m|\le 0.6$ & $ct\ge 0$ \\
$\mu\mu$ & $1.8445\le$$m_{\mu\mu}$$\le 1.8845$ &
$|\delta m|\le 0.6$ & - \\ 
$e\mu$ &  
$1.8445\le$$m_{e\mu}$$\le1.8845$ & 
$|\delta m|\le 0.5$ & - \\\hline\hline
\end{tabular}
\caption{The summary of the optimized event selection criteria
of $D^0\to\ell^+\ell^-$.}
\label{table:sel}
\end{center}
\end{table}
The proper time requirement is found to be useful only 
for the $ee$ mode. 
The background in the $ee$ mode is dominated by combinatorials
with zero average lifetime and is halved, with a reduction of 
less than 20\,\% in signal efficiency, by requiring that the
proper time of the \Dz candidate be positive. Such a requirement
is not applied to the $\mu\mu$ mode, where its reflection 
background has large contribution and a similar lifetime behavior
to the \Dz meson, or to the $e\mu$ mode, where the background is 
very small.

As an important check of the background estimate, we have
compared the expected distribution in the low mass sideband 
( $1.6545\le m_{\ell\ell}\le 1.8045\,\gevcc$ ) with the data.
The peaking background in the low mass sideband is evaluated
from the $\Dz\to h^+h^-$ MC sample using precise measurement 
of the lepton misidentification probabilities.
The random lepton pairs are inferred from the
events in the high mass sideband. 
Unlike the upper side, the combinatoric
background in the low mass
sideband has contributions from the combination of 
two hadrons and the combination of one real lepton with one hadron, 
where the 
hadrons are misidentified as leptons. We estimate those backgrounds 
using 
MC data and known lepton misidentification rates. 
We find that the predicted background distributions and levels 
in the low mass
sideband (before and after the 
optimization of our event selection criteria) have excellent
agreement with our observation in the data for all three
decay modes.

The number of $\Dz\to\pi^+\pi^-$ candidates in the
data, $N_{\pi\pi}$, is extracted by fitting their invariant mass 
distribution 
with a binned maximum likelihood fit. The signal distribution is modeled
as a double Gaussian, and the background distribution is approximated 
as a linear function. The number of reconstructed \Dz mesons
is found to be between 7000 and 12000, depending on the selection
criteria. The relative uncertainties in $N_{\pi\pi}$ are about 1\,\%.

\begin{figure}[htb]
\begin{center}
\scalebox{1}{\includegraphics{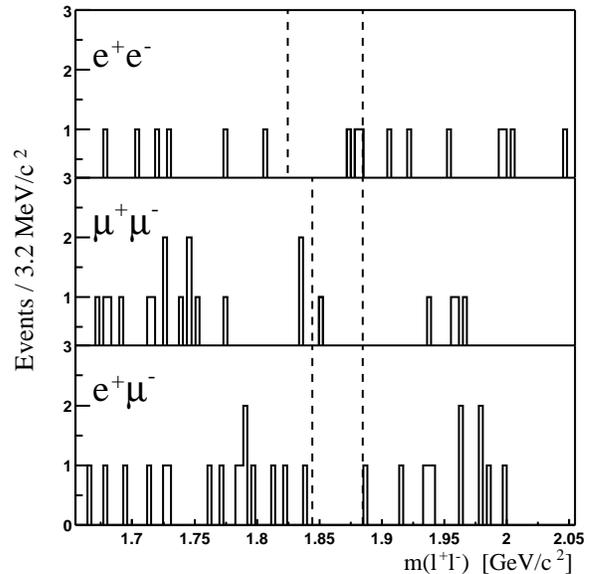}}
\caption{\label{fig:mll}
The dilepton invariant mass distribution for each decay mode.
The dashed lines indicate the optimized signal mass window.}
\end{center}
\end{figure}

The invariant mass distribution of the dilepton candidates 
after applying the optimized event selection 
criteria is shown in Fig.~\ref{fig:mll}. The number of events 
observed ($N_{\rm obs}$) and the expected background ($N_{\rm bg}$) are shown
in Table~\ref{tbl:br}, with no significant excess found in any 
decay mode. 

The largest systematic uncertainty in the
signal efficiency ratio $\epsilon_{\pi\pi}/\epsilon_{\ell\ell}$ 
calculation is due to 
the PID efficiency. It ranges from
1.2\,\% for the $ee$ mode to 4.2\,\% for the $\mu\mu$ mode relative to 
their efficiency ratio.
Other sources of systematic uncertainty are found to be small, including
the track reconstruction efficiencies, track momentum resolution
and MC statistics.

The systematic uncertainties of the background estimate arise 
predominantly from the finite data available in the high mass sideband
for the $ee$ and $e\mu$ modes. 
For the $\mu\mu$ mode, a large fraction of the background is
produced by misidentified $\Dz\to\pip\pim$ decays. The 
relative uncertainty associated with the estimate of muon 
misidentification is found to be about $4.7\,\%$.

\begin{table}[htb]
\begin{center}
\begin{tabular}{lccc}\hline\hline
 & $\Dz\to e^+e^-$ & $\Dz\to \mu^+\mu^-$ & $\Dz\to e^\pm\mu^\mp$ 
\\\hline 
$N^{hh}_{\rm bg}$ & $0.02$ & $3.34\pm0.31$ & $0.21$ \\
$N^{\rm comb}_{\rm bg}$ & $2.21\pm 0.38$ & $1.28\pm0.32$ & $1.93\pm0.36$ \\
$N_{\rm bg}$ & $2.23\pm 0.38$ & $4.63\pm0.45$ & $2.14\pm0.36$ \\
$S\;\;\;[10^{-7}]$ & $2.25\pm0.12$ & $4.53\pm0.30$ 
& $3.27\pm0.20$\\\hline
$N_{\rm obs}$ & 3 & 1 & 0 \\
UL obtained & $1.2\times10^{-6}$ & $1.3\times10^{-6}$ 
& $8.1\times 10^{-7}$ \\\hline\hline
\end{tabular}
\caption{The summary of the number of expected background
events ($N_{\rm bg}$), the sensitivity factor ($S$), number of 
observed events ($N_{\rm obs}$), 
and the branching fraction upper limits at the 90\,\% 
confidence level for each decay modes.
The uncertainties quoted here are total uncertainties. 
The uncertainty of $N^{hh}_{\rm bg}$ is negligible 
for the $ee$ and $e\mu$ decay modes.}
\label{tbl:br}
\end{center}
\end{table}

The branching fraction upper limits (UL)
have been calculated including 
all uncertainties using an extended version~\cite{Conrad:2002kn} 
of the Feldman-Cousins method~\cite{Feldman:1997qc}. All of the uncertainties
have a negligible effect on the limits. The results 
are listed in Table~\ref{tbl:br}.

In summary, we have performed a search for the FCNC decays
 $\Dz\to e^+e^-$,
$\Dz\to \mu^+\mu^-$, and the LFV decays $\Dz\to e^\pm\mu^\mp$ 
using the
\babar\ detector. No evidence is found for these decays. 
The upper
limits on the branching fractions at the 90\,\% confidence level
are  $1.2\times 10^{-6}$ for
$\Dz\to e^+e^-$, $1.3\times 10^{-6}$ for $\Dz\to\mu^+\mu^-$, and
$8.1\times 10^{-7}$ for
$\Dz\to e^\pm\mu^\mp$.
Our result improves the present best limits
by a factor of 5 for the $ee$ mode~\cite{Aitala:1999db}, 
a little less than 2 for
the $\mu\mu$ mode~\cite{Abt:2004hn}, and 
10 for the $e\mu$ mode~\cite{Aitala:1999db}.
The upper limits for the branching fractions of the $e\mu$ 
and $\mu\mu$ modes
begin to confine the allowed parameter space of R-parity violating
supersymmetric models~\cite{Burdman:2001tf}.

We are grateful for the excellent luminosity and machine conditions
provided by our \pep2\ colleagues, 
and for the substantial dedicated effort from
the computing organizations that support \babar.
The collaborating institutions wish to thank 
SLAC for its support and kind hospitality. 
This work is supported by
DOE
and NSF (USA),
NSERC (Canada),
IHEP (China),
CEA and
CNRS-IN2P3
(France),
BMBF and DFG
(Germany),
INFN (Italy),
FOM (The Netherlands),
NFR (Norway),
MIST (Russia), and
PPARC (United Kingdom). 
Individuals have received support from CONACyT (Mexico), A.~P.~Sloan Foundation, 
Research Corporation,
and Alexander von Humboldt Foundation.


\begin{thebibliography}{99}

\bibitem{conjugate}
Throughout this paper charge conjugation
is implied.

\bibitem{Glashow:gm}
S.~L.~Glashow, J.~Iliopoulos and L.~Maiani,
Phys.\ Rev.\ D {\bf 2}, 1285 (1970).

\bibitem{neutrino}
This process is allowed within extensions  
to the SM that have non-zero neutrino mass. 

\bibitem{Burdman:2001tf}
G.~Burdman, E.~Golowich, J.~Hewett and S.~Pakvasa,
Phys.\ Rev.\ D {\bf 66}, 014009 (2002).

\bibitem{Aubert:2001tu}
B.~Aubert {\it et al.}  [BABAR Collaboration],
Nucl.\ Instrum.\ Meth.\ A {\bf 479}, 1 (2002).

\bibitem{Hagiwara:fs}
K.~Hagiwara {\it et al.}  [Particle Data Group Collaboration],
Phys.\ Rev.\ D {\bf 66}, 010001 (2002).

\bibitem{Sjostrand:2000wi}
T.~Sjostrand, P.~Eden, C.~Friberg, L.~Lonnblad, G.~Miu, S.~Mrenna and E.~Norrbin,
Comput.\ Phys.\ Commun.\  {\bf 135}, 238 (2001).

\bibitem{Barberio:1993qi}
E.~Barberio and Z.~Was,
Comput.\ Phys.\ Commun.\  {\bf 79}, 291 (1994).

\bibitem{Agostinelli:2002hh}
S.~Agostinelli {\it et al.}  [GEANT4 Collaboration],
Nucl.\ Instrum.\ Meth.\ A {\bf 506}, 250 (2003).

\bibitem{Feldman:1997qc}
G.~J.~Feldman and R.~D.~Cousins,
Phys.\ Rev.\ D {\bf 57}, 3873 (1998).

\bibitem{Conrad:2002kn}
J.~Conrad, O.~Botner, A.~Hallgren and C.~Perez de los Heros,
Phys.\ Rev.\ D {\bf 67}, 012002 (2003).

\bibitem{Aitala:1999db}
E.~M.~Aitala {\it et al.}  [E791 Collaboration],
Phys.\ Lett.\ B {\bf 462}, 401 (1999).


\bibitem{Abt:2004hn}
I.~Abt {\it et al.}  [HERA-B Collaboration],
Phys.\ Lett.\ B {\bf 596}, 173 (2004).


\end{thebibliography}
\end{document}